\documentclass[a4paper,conference]{IEEEtran}

\usepackage{amssymb}
\usepackage{amsfonts}
\usepackage[colorlinks,citecolor=blue,urlcolor=blue,breaklinks=true]{hyperref}
\usepackage{tikz}
\usepackage{ifthen}
\usepackage[cmex10]{amsmath} 
\usepackage{arydshln}

\begin{document}
\title{Information Theoretic Authentication and Secrecy Codes in the Splitting Model}

\author{%
\IEEEauthorblockN{Michael Huber}
\IEEEauthorblockA{University of Tuebingen\\
              Wilhelm Schickard Institute for Computer Science\\
              Sand~13, D-72076 Tuebingen, Germany\\
              Email: michael.huber@uni-tuebingen.de}
}

\newtheorem{proposition}{Proposition}
\newtheorem{definition}{Definition}
\newtheorem{corollary}{Corollary}
\newtheorem{theorem}{Theorem}
\newtheorem{lemma}{Lemma}
\newtheorem{problem}{Problem}
\newtheorem{example}{Example}
\newtheorem{remark}{Remark}

\maketitle

\begin{abstract}
In the splitting model, information theoretic authentication codes allow non-deterministic encoding, that is, several messages can be used to communicate a particular plaintext. \linebreak Certain applications require that the aspect of secrecy should hold simultaneously.   
Ogata--Kurosawa--Stinson--Saido (2004) have constructed optimal splitting authentication codes achieving perfect secrecy for the special case when the number of keys equals the number of messages. In this paper, we establish a construction method for optimal splitting authentication codes with perfect secrecy in the more general case when the number of keys may differ from the number of messages. To the best knowledge, this is the first result of this type.
\end{abstract}

\section{Introduction}\label{Intro}

The development of quantum computer resistant cryptographic schemes and security technologies is of crucial importance for maintaining cryptographic long-term security and/or confidentiality of digital data, as classical cryptographic primitives such as RSA, DSA, or ECC would be easily broken by future quantum computing based attacks (e.g.,~\cite{Buch06,bern09}). Application areas where cryptographic long-term security and/or confidentiality is strongly required include archiving official documents, notarial contracts, court records, medical data, state secrets, copyright protection as well as further areas concerning e-government, e-health, e-publication, et cetera.

To this end, one promising approach is the design of information theoretic authentication and secrecy systems (e.g.,~\cite{mau99,Hu2010}). The information theoretic, or unconditional, security model does not depend on any complexity assumptions and hence cannot be broken given unlimited computational resources. This guarantees not only resistance against future quantum computing based attacks but also perfect security in the classical world.

This paper considers authentication and secrecy codes in the splitting model. Splitting is of importance, for instance, in the context of authentication with arbitration~\cite{Kur01}  (i.e., protection against insider attacks in addition to outsider attacks). 
Ogata--Kurosawa--Stinson--Saido~\cite{Ogata04}  have constructed optimal splitting authentication codes with perfect secrecy for the special case when the number of keys equals the number of messages. In this work, we develop a construction method for optimal splitting authentication codes with perfect secrecy in the more general case when the number of keys may differ from the number of messages. To the best knowledge, this is the first result of this type.
Our simple yet powerful approach is based on the notion of \emph{cyclic} splitting designs and establishes an efficient method to construct optimal splitting authentication codes with perfect secrecy.  

\section{The Splitting Model}\label{SplittModel}

We rely on the information theoretical, or unconditional secure, authentication model developed by Simmons (e.g.,~\cite{Sim82,Sim92}). Our notation follows~\cite{Ogata04,Stin90,Hu2010b}.
In this model, three participants are involved: a \emph{transmitter}, a \emph{receiver}, and an \emph{opponent}.  The transmitter wants to communicate information to the receiver via a public communications channel. The receiver in return would like to be confident that any received information actually came from the transmitter and not from some opponent (\emph{integrity} of information). The transmitter and the receiver are assumed to trust each other. An authentication code is sometimes called, for short,  an \emph{$A$-code}.

Let $\mathcal{S}$ denote a finite set of \emph{source states} (or \emph{plaintexts}), $\mathcal{M}$ a finite set of \emph{messages} (or \emph{ciphertexts}), and $\mathcal{E}$ a finite set of \emph{encoding rules} (or \emph{keys}). Using an encoding rule $e\in \mathcal{E}$,
the transmitter encrypts a source state $s \in \mathcal{S}$ to obtain the message $m=e(s)$ to be sent over the channel.
The encoding rule is communicated to the receiver via a secure channel prior to any messages being sent.
When it is possible that more than one message can be used to communicate a particular source state $s \in \mathcal{S}$ under the same encoding rule $e\in \mathcal{E}$, then the authentication code is said to have \emph{splitting}. In this case, a message $m \in \mathcal{M}$ is computed as $m=e(s,r)$, where $r$ denotes a random number chosen from some specified finite set $\mathcal{R}$. If we define
\[e(s):=\{m \in \mathcal{M}: m = e(s,r)\;\, \mbox{for some}\;\, r \in \mathcal{R}\}\]
for each encoding rule $e\in \mathcal{E}$ and each source state $s \in \mathcal{S}$,
then splitting means that $\left|e(s)\right|>1$ for some $e\in \mathcal{E}$ and some $s \in \mathcal{S}$.
In order to ensure that the receiver can decrypt the message being sent, it is required for any $e \in \mathcal{E}$ that $e(s) \cap e(s^\prime)=\emptyset$ if $s \neq s^\prime$.
For a given encoding rule $e \in \mathcal{E}$, let \[M(e):= \bigcup_{s \in \mathcal{S}} e(s)\] denote the set of \emph{valid} messages. A received message $m$ will be accepted by the receiver as being authentic if and only if $m \in M(e)$. When this is fulfilled, the receiver decrypts the message $m$ by applying the decoding rule $e^{-1}$, where \[e^{-1}(m)=s\;\, \mbox{if}\;\,  m = e(s,r)\;\, \mbox{for some}\;\, r \in \mathcal{R}.\]
A splitting authentication code is called \emph{$c$-splitting} if
\[\left|e(s)\right|=c\]
for every encoding rule $e \in \mathcal{E}$ and every source state \mbox{$s \in \mathcal{S}$}.
We note that an authentication code can be represented algebraically by a $\left|\mathcal{E}\right| \times \left|\mathcal{S}\right|$ \emph{encoding matrix} with the rows indexed by the encoding rules $e \in \mathcal{E}$, the columns indexed by the source states $s \in \mathcal{S}$, and the entries defined by $a_{es}:=e(s)$. 

We address the scenario of a \emph{spoofing attack} of order $i$ (cf.~\cite{Mass86}):
Suppose that an opponent observes $i\geq 0$ distinct messages, which are sent through the public channel using the same encoding rule. The opponent then inserts a new message $m^\prime$ (being distinct from the $i$ messages already sent), hoping to have it accepted by the receiver as authentic. The cases $i=0$ and $i=1$ are called \emph{impersonation game} and \emph{substitution game}, respectively. 

For any $i$, we assume that there is some probability distribution on the set of \mbox{$i$-subsets} of source states, so that any set of $i$ source states has a non-zero probability of occurring. For simplification, we ignore the order in which the $i$ source states occur, and assume that no source state occurs more than once.
Given this probability distribution on the set $\mathcal{S}$ of source states, the receiver and transmitter also choose a probability distribution on the set $\mathcal{E}$ of encoding rules, called an \emph{encoding strategy}. It is assumed that the opponent knows the encoding strategy being used. If splitting occurs, then the receiver/transmitter will also choose a \emph{splitting strategy} to determine $m \in \mathcal{M}$, given $s \in \mathcal{S}$ and $e \in \mathcal{E}$ (this corresponds to \emph{non-deterministic encoding}). The transmitter/receiver will determine these strategies to minimize the chance of being deceived by the opponent.
The \emph{deception probability} $P_{d_i}$ denotes the probability that the opponent can deceive the transmitter/receiver with a spoofing attack of order $i$.

\section{Combinatorial Splitting Designs}\label{SplittDesigns}

The notion of splitting balanced incomplete block designs and, more generally, that of splitting $t$-designs have been introduced in~\cite{Ogata04} and~\cite{Hu2010b}, respectively.

\begin{definition}\label{splittdesign}
For positive integers $t,v,b,c,u,\lambda$ with \mbox{$t \leq u$} and \mbox{$cu \leq v$}, a
\mbox{$t$-$(v,b,l=cu,\lambda)$} \emph{splitting design} $\mathcal{D}$ is a pair \mbox{$(X,\mathcal{B})$}, satisfying
the following properties:

\begin{enumerate}

\item[(i)] $X$ is a set of $v$ elements, called \emph{points},

\smallskip

\item[(ii)] $\mathcal{B}$ is a family of \mbox{$l$-subsets} of $X$, called \emph{blocks}, such that every block $B_i \in \mathcal{B}$
$(1\leq i \leq \left|\mathcal{B}\right|=:b)$ is expressed as a disjoint union \[B_i= B_{i,1} \cup \cdots \cup B_{i,u}\] with $\left|B_{i,1}\right| = \cdots = \left|B_{i,u}\right|=c$ and $\left|B_i\right|=l=cu$,

\smallskip

\item[(iii)]  every \mbox{$t$-subset} $\{x_m\}_{m=1}^t $of $X$ is contained in exactly $\lambda$ blocks $B_i=B_{i,1} \cup \cdots \cup B_{i,u}$ such that
\[x_m \in B_{i,j_m} \quad (j_m \; \mbox{between} \;  1 \; \mbox{and} \; u)\]
for each $1\leq m \leq t$,  and $j_1, \ldots, j_t$ are mutually distinct.
\end{enumerate}
\end{definition}

We summarize some basic conditions concerning the existence of splitting designs (cf.~\cite{Ogata04,Hu2010b}).

\begin{proposition}\label{s-design_splitt}
Let $\mathcal{D}=(X,\mathcal{B})$ be a \mbox{$t$-$(v,b,l=cu,\lambda)$} splitting design, and for a positive integer $s \leq t$, let $S \subseteq X$
with $\left|S\right|=s$. Then the number of blocks containing
each element of $S$ as per Definition~\ref{splittdesign} is given by
\[\lambda_s = \lambda \frac{{v-s \choose t-s}}{c^{t-s}{u-s \choose t-s}}.\]
In particular, for $t\geq 2$, a \mbox{$t$-$(v,b,l=cu,\lambda)$} splitting design is
also an \mbox{$s$-$(v,b,l=cu,\lambda_s)$} splitting design.
\end{proposition}

\begin{proposition}\label{Comb_t=5_splitt}
Let $\mathcal{D}=(X,\mathcal{B})$ be a \mbox{$t$-$(v,b,l=cu,\lambda)$} splitting design. Let $r:= \lambda_1$ denote the
number of blocks containing a given point. Then the following holds:

\begin{enumerate}

\item[{(a)}] $bl = vr.$

\smallskip

\item[{(b)}] $\displaystyle{{v \choose t} \lambda = bc^t {u \choose t}.}$

\smallskip

\item[{(c)}] $rc^{t-1}(u-1)=\lambda_2(v-1)$ for $t \geq 2$.

\end{enumerate}
\end{proposition}

\begin{proposition}\label{divCond}
Let $\mathcal{D}=(X,\mathcal{B})$ be a \mbox{$t$-$(v,b,l=cu,\lambda)$} splitting design.
Then
\[\lambda {v-s \choose t-s} \equiv  0 \, \bigg(\hspace{-0.3cm} \mod c^{t-s}{u-s \choose t-s}\bigg)\]
for each positive integer $s \leq t$.
\end{proposition}

\begin{proposition}\label{FisherIn_splitt}
If $\mathcal{D}=(X,\mathcal{B})$ is a \mbox{$t$-$(v,b,l=cu,\lambda)$} splitting design with $t \geq 2$, then
\[b\geq \frac{v}{u}.\]
\end{proposition}

\section{Optimal Splitting Authentication Codes}\label{Auth}

We state lower bounds on cheating probabilities for splitting authentication codes (cf.~\cite{DeSoete91,Blund99}).

\begin{theorem}\label{deceptprob_splitt}
In a splitting authentication code, for every $0 \leq i \leq t$, the deception probabilities are bounded below by
\[P_{d_i}\geq \min_{e \in \mathcal{E}} \frac{\left|M(e)\right| - i \cdot \max_{s \in \mathcal{S}} \left| e(s) \right|}{\left|\mathcal{M}\right|-i}.\]
\end{theorem}

A splitting authentication code is called $t$\emph{-fold secure against spoofing} if
\[P_{d_i}= \min_{e \in \mathcal{E}} \frac{\left|M(e)\right| - i \cdot \max_{s \in \mathcal{S}} \left| e(s) \right|}{\left|\mathcal{M}\right|-i}\]
for all $0 \leq i \leq t$.

\smallskip

We indicate a lower bound on the size of encoding rules for splitting authentication codes (see~\cite{Hu2010b}, and~\cite{Brick84,Sim90} for the case $t=2$).

\begin{theorem}\label{my1}
If a splitting authentication code is \mbox{$(t-1)$-fold} secure against spoofing, then the number of encoding rules is bounded below by
\[\left|\mathcal{E}\right|  \geq \prod_{i=0}^{t-1} \frac{\left|\mathcal{M}\right|  -i}{\left|M(e)\right| - i \cdot \max_{s \in \mathcal{S}}\left| e(s) \right|}.\]
\end{theorem}

A splitting authentication code is called \emph{optimal} if the number of encoding rules meets the lower bound with equality.

\begin{corollary}\label{Cor1}
In a $c$-splitting authentication code,
\[P_{d_i}\geq \frac{c(\left|\mathcal{S}\right|-i)}{\left|\mathcal{M}\right|-i}\]
for every $0 \leq i \leq t$.
\end{corollary}

\begin{corollary}\label{Cor2}
If a $c$-splitting authentication code is \mbox{$(t-1)$}-fold secure against spoofing, then
\[\left|\mathcal{E}\right|  \geq \frac{{\left|\mathcal{M}\right| \choose t}}{c^t{\left|\mathcal{S}\right| \choose t}}.\]
\end{corollary}

Optimal splitting authentication codes can be characterized in terms of splitting designs (see~\cite{Hu2010b}, and~\cite{Ogata04} for the case $t=2$) as follows.

\begin{theorem}\label{my2}
Suppose there is a \mbox{$t$-$(v,b,l=cu,1)$} splitting design with $t \geq 2$. Then there is an optimal \mbox{$c$-splitting} authentication code for $u$ equiprobable source states, having $v$ messages and ${v \choose t}/[c^t{u \choose t}]$ encoding rules, that is $(t-1)$-fold secure against spoofing. Conversely, if there is an optimal \mbox{$c$-splitting} authentication code for $u$ source states, having  $v$ messages and ${v \choose t}/[c^t{u \choose t}]$ encoding rules, that is \mbox{$(t-1)$-fold} secure against spoofing, then there is a \mbox{$t$-$(v,b,l=cu,1)$} splitting design.
\end{theorem}

\section{Optimal Splitting Authentication Codes with Perfect Secrecy}\label{New}

In what follows, we are interested in optimal splitting authentication codes that simultaneously achieve perfect secrecy. According to Shannon~\cite{Shan49},
an authentication code is said to have \emph{perfect secrecy} if
\[p_S(s | m)=p_S(s)\]
for every source state $s \in \mathcal{S}$ and every message $m \in \mathcal{M}$, that is, the \emph{a posteriori} probability that the source state is $s$, given that the message $m$ is observed, is identical to
the \emph{a priori} probability that the source state is $s$.

By introducing the notion of an external difference family (EDF) (which yields a certain type of a splitting design), Ogata--Kurosawa--Stinson--Saido~\cite[Thm.\,3.4]{Ogata04}  have given a construction scheme for optimal splitting authentication codes with perfect secrecy in the special case when the number of keys equals the number of messages.

\begin{theorem}\label{ogata04perfsecr}
Suppose there exists a \mbox{$(v,c,1)$} $u$-EDF over an Abelian group of order $v$, then there is an optimal \mbox{$c$-splitting} authentication code for $u$ equiprobable source states, having $v$ messages and $v$ encoding rules, that is one-fold secure against spoofing and simultaneously achieves perfect secrecy. 
\end{theorem}

An example is as follows (cf.~\cite[Exs.\,2.3\,\&\,5.2]{Ogata04}).

\begin{example}\label{example_concrete}
An optimal \mbox{$2$-splitting} authentication code for $u=2$ equiprobable source states, having $v=9$ messages and $b=9$ encoding rules, that is one-fold secure against spoofing and achieves perfect secrecy can be constructed from a \mbox{$2$-$(9,9,4=2\times2,1)$} splitting design. Each encoding rule is used with probability $1/9$. An encoding matrix is given in Table~\ref{encod}.

\begin{table}
\renewcommand{\arraystretch}{1.3}
\caption{Splitting authentication code with perfect secrecy from a \mbox{$2$-$(9,9,4=2\times2,1)$} splitting design.}\label{encod}

\begin{center}
\begin{tabular}{|c| c c|}
  \hline
  & $s_1$ & $s_2$ \\
  \hline
  $e_1$ & \{1,2\} & \{3,5\}\\
  $e_2$ & \{2,3\} & \{4,6\}\\
  $e_3$ & \{3,4\} & \{5,7\}\\
  $e_4$ & \{4,5\} & \{6,8\}\\
  $e_5$ & \{5,6\} & \{7,9\}\\
  $e_6$ & \{6,7\} & \{8,1\}\\
  $e_7$ & \{7,8\} & \{9,2\}\\
  $e_8$ & \{8,9\} & \{1,3\}\\
  $e_9$ & \{9,1\} & \{2,4\}\\
  \hline
\end{tabular}
\end{center}
\end{table}

\end{example}

\pagebreak

In the following, we develop a construction method for obtaining optimal splitting authentication codes with perfect secrecy in the more general case when the number of keys may differ from the number of messages:

\begin{enumerate}

\item[(1)] We first introduce the notion of a \emph{cyclic} splitting design.  
Let $\mathcal{D}=(X,\mathcal{B})$ be a \mbox{$2$-$(v,b,l=cu,\lambda)$} splitting design, and let $\sigma$ be a permutation on $X$. For a block $B_i=\{B_{i,1},\ldots,B_{i,u}\} \in \mathcal{B}$ given as in (ii) of Definition~\ref{splittdesign}, define $B_i^\sigma :=\{B_{i,1}^\sigma,\ldots,B_{i,u}^\sigma\}$, satisfying
\[B_i^\sigma= B_{i,1}^\sigma \cup \cdots \cup B_{i,u}^\sigma\] with $\left|B_{i,1}^\sigma\right| = \cdots = \left|B_{i,u}^\sigma\right|=c$ and $\left|B_i^\sigma\right|=l=cu$.
If $\mathcal{B}^\sigma:=\{B_i^\sigma : B_i \in \mathcal{B}, 1 \leq i \leq b\}=\mathcal{B}$, then $\sigma$ is called an \emph{automorphism} of $\mathcal{D}$. If there exists an automorphism $\sigma$ of order $v$, then $\mathcal{D}$ is called \emph{cyclic}. In this case, the point-set $X$ can be identified with $\mathbb{Z}_v$, the set of integers modulo $v$, and $\sigma$ can be represented by $\sigma : j \rightarrow j + 1$ (mod $v$).
For a block $B_i=\{B_{i,1},\ldots,B_{i,u}\}$,  the set 
\[B_i + j := \{B_{i,1}+j \;(\mbox{mod}\; v), \ldots, B_{i,u}+j \; (\mbox{mod}\; v)\}\] for $j \in \mathbb{Z}_v$ is called a \emph{translate} of $B_i$, and the set of all distinct translates of $B_i$ is called the \emph{orbit} containing $B_i$. If the length of an orbit is $v$, then the orbit is said to be \emph{full}, otherwise \emph{short}. A block chosen arbitrarily from an orbit is called a \emph{base block} (or \emph{starter block}). For a cyclic \mbox{$2$-$(v,b,l=cu,1)$} splitting design to exist, a necessary condition is $v \equiv 1$ or $l$ $($mod $u(u-1)c^2)$. When $v \equiv 1$ $($mod $u(u-1)c^2)$ all orbits are full.

\item[(2)] Let us assume that there exists a cyclic \mbox{$2$-$(v,b,l=cu,1)$} splitting design without short orbit. Then, by Theorem~\ref{my2}, there is  an optimal \mbox{$c$-splitting} authentication code for $u$ equiprobable source states, having $v$ messages and ${v \choose 2}/[c^2{u \choose 2}]$ encoding rules, that is one-fold secure against spoofing. Furthermore,  when considering the corresponding $b \times u$ encoding matrix, it follows by constructional reasons from the underlying cyclic splitting design without short orbit that the code simultaneously achieves perfect secrecy under the assumption that the encoding rules are used with equal probability.

\end{enumerate}
 
Hence, we have proved the following theorem.

\begin{theorem}\label{my3}
Suppose there is a cyclic \mbox{$2$-$(v,b,l=cu,1)$} splitting design without short orbit (that is, it holds that $v \equiv 1$ $($mod $u(u-1)c^2)$). 
Then there is an optimal \mbox{$c$-splitting} authentication code for $u$ equiprobable source states, having $v$ messages and ${v \choose 2}/[c^2{u \choose 2}]$ encoding rules, that is one-fold secure against spoofing and simultaneously achieves perfect secrecy. 
\end{theorem}

Relying on some recent constructions of splitting designs (cf.~\cite[Sect.\,3.2]{Ge05}), we give exemplarily a series of optimal splitting authentication codes with perfect secrecy.

\begin{example}\label{example_series}

\begin{enumerate}

\item[(i)] An optimal \mbox{$2$-splitting} authentication code for $u=2$ equiprobable source states, having $v=17$ messages and $b=34$ encoding rules, that is one-fold secure against spoofing and achieves perfect secrecy can be constructed from a cyclic \mbox{$2$-$(17,34,4=2\times2,1)$} splitting design with base blocks $\{\{1,2\},\{3,5\}\}$ and $\{\{1,2\},\{11,13\}\}$. Each encoding rule is used with probability $1/34$. An encoding matrix is given in Table~\ref{encod2}.

\begin{table}
\renewcommand{\arraystretch}{1.3}
\caption{Splitting authentication code with perfect secrecy from a cyclic \mbox{$2$-$(17,34,4=2\times2,1)$} splitting design with base blocks $\{\{1,2\},\{3,5\}\}$ and $\{\{1,2\},\{11,13\}\}$.}\label{encod2}

\begin{center}
\begin{tabular}{|c| c c|}
  \hline
  & $s_1$ & $s_2$ \\
  \hline
  $e_1$ & \{1,2\} & \{3,5\}\\
  $e_2$ & \{2,3\} & \{4,6\}\\
  $e_3$ & \{3,4\} & \{5,7\}\\
  $e_4$ & \{4,5\} & \{6,8\}\\
  $e_5$ & \{5,6\} & \{7,9\}\\
  $e_6$ & \{6,7\} & \{8,10\}\\
  $e_7$ & \{7,8\} & \{9,11\}\\
  $e_8$ & \{8,9\} & \{10,12\}\\
  $e_9$ & \{9,10\} & \{11,13\}\\
  $e_{10}$ & \{10,11\} & \{12,14\}\\
  $e_{11}$ & \{11,12\} & \{13,15\}\\
  $e_{12}$ & \{12,13\} & \{14,16\}\\
  $e_{13}$ & \{13,14\} & \{15,17\}\\
  $e_{14}$ & \{14,15\} & \{16,1\}\\
  $e_{15}$ & \{15,16\} & \{17,2\}\\
  $e_{16}$ & \{16,17\} & \{1,3\}\\
  $e_{17}$ & \{17,1\} & \{2,4\}\\ 
  \hdashline  
  $e_{18}$ & \{1,2\} & \{11,13\}\\   
  $e_{19}$ & \{2,3\} & \{12,14\}\\    
  $e_{20}$ & \{3,4\} & \{13,15\}\\    
  $e_{21}$ & \{4,5\} & \{14,16\}\\  
  $e_{22}$ & \{5,6\} & \{15,17\}\\
  $e_{23}$ & \{6,7\} & \{16,1\}\\
  $e_{24}$ & \{7,8\} & \{17,2\}\\
  $e_{25}$ & \{8,9\} & \{1,3\}\\
  $e_{26}$ & \{9,10\} & \{2,4\}\\
  $e_{27}$ & \{10,11\} & \{3,5\}\\
  $e_{28}$ & \{11,12\} & \{4,6\}\\
  $e_{29}$ & \{12,13\} & \{5,7\}\\   
  $e_{30}$ & \{13,14\} & \{6,8\}\\   
  $e_{31}$ & \{14,15\} & \{7,9\}\\    
  $e_{32}$ & \{15,16\} & \{8,10\}\\    
  $e_{33}$ & \{16,17\} & \{9,11\}\\  
  $e_{34}$ & \{17,1\} & \{10,12\}\\       
  \hline
\end{tabular}
\vspace*{-0.57cm}
\end{center}
\end{table}

\item[(ii)] As generalization of (i), an optimal \mbox{$c$-splitting} authentication code for $u=2$ equiprobable source states, having $v=2c^2n+1$ messages and $b=(2c^2n+1)n$ encoding rules, that is one-fold secure against spoofing and achieves perfect secrecy can be constructed from a cyclic \mbox{$2$-$(2c^2n+1,(2c^2n+1)n,l=c\times2,1)$} splitting design with base blocks $\{\{1,2,\ldots,c\},\{2c^2h-(2c^2-c)+1, 2c^2h-(2c^2-c)+c+1,\ldots, 2c^2h-(2c^2-c)+c(c-1)+1\}\}$ for all $1\leq h \leq n$ .

\item[(iii)] Further examples of splitting authentication codes with perfect secrecy, also for $u>2$, can be obtained in the same way from various further constructions of splitting designs in~\cite[Sect.\,3.2]{Ge05}.
\end{enumerate}

\end{example}

\section*{Acknowledgment}

The author gratefully acknowledges support by the Deutsche Forschungsgemeinschaft (DFG) via a Heisenberg grant (Hu954/4) and a Heinz Maier-Leibnitz Prize grant (Hu954/5).

\end{document}